\documentclass[12pt]{article}
\usepackage{graphicx}
\newcommand{\lsim}{\lower.7ex\hbox{$\;\stackrel{\textstyle<}{\sim}\;$}}

\begin{document}
\def\cw{c_{\omega}}
\def\sw{s_{\omega}}
\def\m1{m_{h}}
\def\m2{m_{H}}
\def\tev{\, {\rm TeV}}
\def\gev{\, {\rm GeV}}
\def\comment#1{{\bf [#1]}}
\def\beq{\begin{equation}}
\def\eeq{\end{equation}}
\def\slashET{{E_T\!\!\!\!\!\!/\,\,\,}}
\def\xfb{\,{\rm fb}}

\title{
\begin{flushright}
{\rm {\normalsize{MCTP-07-02}}}\\
\vspace{1cm}
\end{flushright}
Narrow Trans-TeV Higgs Bosons and $H\rightarrow hh$ Decays: \\
Two LHC Search Paths for a Hidden Sector Higgs Boson}

\author{Matthew Bowen$^{1,2}$, Yanou Cui$^2$, James D. Wells$^2$\\[.3cm]
$^1${\small{\it Department of Physics, University of Washington,}}
{\small{\it Seattle, WA 98195, USA}}
\\
$^2${\small{\it Michigan Center For Theoretical Physics (MCTP)}}\\
{\small{\it Department of Physics, University of Michigan,}}
             {\small {\it Ann Arbor, MI 48109, USA}} \\
 }

\maketitle
\begin{abstract}

We consider the addition of a condensing singlet scalar field to the
Standard Model. Such a scenario may be motivated by any
number of theoretical ideas, including the common result in string-inspired
model building of
singlet scalar fields charged under some hidden sector gauge symmetry.  For
concreteness, we specify an example model of this type, and
consider the relevant constraints on Higgs physics, such as
triviality, perturbative unitarity and precision electroweak
analysis.  We then show that
there are two unique features of the phenomenology that present
opportunities for discovery at the Large Hadron Collider (LHC). First, it is possible to
identify and discover a narrow trans-TeV Higgs boson in this
scenario --- a mass scale that is well above the scale at which it
is meaningful to discuss a SM Higgs boson. Second, the decays of
the heavier scalar state into the lighter Higgs bosons can proceed
at a high rate and may be the first discovery mode in the Higgs
sector.

\end{abstract}

\setcounter{equation}{0}

\vfill\eject

\tableofcontents

\section{Introduction}
\indent

Nature may contain many more particles than are implied by what we
consider at first thought to be well-motivated ideas of new physics.  This is especially
true if our mindset is entirely on trying to understand
electroweak symmetry breaking in the most minimal framework that
we can devise.  There are many things to explain in nature beyond the Standard
Model (SM), and electroweak symmetry breaking
(EWSB) is merely one of them.

In the search for beyond Standard Model (SM) physics,
hidden sector SM singlet fields  are ubiquitous in many existing
theories, such as in string-inspired particle physics models
containing many more gauge groups and corresponding scalar sectors
than would otherwise be needed to describe or contain the SM.  Even
without these more practical motivations, it is always reasonable
to imagine such a `phantom' world and how it can influence the
physics of our SM world, since no present experimental data rules
out its existence, and we know so little definitively about the
Higgs sector.

It would be interesting to pursue whether we can find
evidence for the hidden sector at the LHC. In many scenarios,
hidden sector fields couple to the SM fields only through
non-renormalizable terms or loop effects. In these cases, the
discovery of these fields seems not very promising and they may
end up to be truly `hidden' at colliders. Fortunately, there are
two renormalizable interactions between the hidden sector
and the SM fields. The first one is the mixing between $U(1)_Y$
and the $U(1)_{hid}$ through the kinetic term $\chi
B_{\mu\nu}C^{\mu\nu}$ where $B,C$ are the field strengths of the
two Abelian fields, respectively. This consideration leads to
$Z'$ physics, which has been well studied\cite{Z-prime}. In this
paper we will focus on the phenomenology of the other possibility
which applies to more general hidden gauge structure (not just
Abelian groups): the renormalizable interaction of the SM Higgs
with the hidden sector Higgs.  There are few ways
that the SM fields can interact with the hidden sector or phantom sector
fields, and the Higgs boson, which can form a gauge-invariant
dimension-2 operator all on its own, is a prime candidate to
pursue this connection~\cite{Schabinger:2005ei,Patt:2006fw,Strassler,others,related}.

  Concretely, the analysis in this paper is based on the model
presented in \cite{Schabinger:2005ei}, where the SM Higgs $\Phi_{SM}$ couples to a
hidden scalar $\Phi_H$ through the renormalizable term
$|\Phi_{SM}|^2|\Phi_H|^2$. We
also assume that the hidden sector has a rich gauge theory
structure which is at least partly broken by
$\langle\Phi_H\rangle\neq0$. A nontrivial
vev of $\Phi_H$ is necessary for the mass mixing between the SM
Higgs and $\Phi_H$, which results in two mass eigenstates,
$h$, $H$. It is this mixing that brings in the
two possible distinct signatures at the LHC which are of primary
interest in this paper: a narrow width trans-TeV Higgs boson and the
observable $H\rightarrow hh$ decay.

    Here is the outline of what follows. In section 2, we give a brief
review of the model we will analyze. In section 3, we study the
bounds on Higgs masses for this model, based on the considerations
of perturbative unitarity, triviality and precision electroweak
measurements. We find that the canonical
constraints on the upper limit of the Higgs mass do not apply for the
heavier Higgs boson $H$ because of the mixing effect. Based on the
results of the earlier sections, we propose two possible intriguing features
to be probed at
future colliders: narrow trans-TeV Higgs boson and $H\rightarrow hh$
decay width. In section 4, we study the LHC implications of those two
signatures in detail and demonstrate that they can be
distinguishable and therefore shed new light on beyond SM physics.

 \section{Model Review}
 \indent

   To be self-contained, we first briefly review the model in~\cite{Schabinger:2005ei},
 which sets the framework and notation for what we analyze here.
 We assume that
 there is a hidden $U(1)$ gauge symmetry which is broken by a
 vacuum expectation value (vev) of the Higgs boson $\Phi_H$. We denote
 the $U(1)_{hid}$ gauge boson as $V$, which gets a mass $m_V$ after the
 breaking of $U(1)_{hid}$. In this model, the hidden sector
 Higgs boson $\Phi_H$ mixes with the SM Higgs $\Phi_{SM}$ through a
 renormalizable interaction $|\Phi_{SM}|^2|\Phi_H|^2$. The Higgs boson
 Lagrangian\footnote{Although we do not discuss it specifically in this work, there is
 an analogous supersymmetric construction where the two Higgs fields interact
 via a $D$-term from a shared $U(1)$ symmetry~\cite{Schabinger:2005ei}.} under consideration is
 \begin{equation}
 \mathcal{L}_{Higgs}=|\mathcal{D}_{\mu}\Phi_{SM}|^2+|\mathcal{D}_{\mu}\Phi_H|^2+
 m^2_{\Phi_{SM}}|\Phi_{SM}|^2+m^2_{\Phi_H}|\Phi_H|^2-\lambda|\Phi_{SM}|^4
 -\rho|\Phi_H|^4-\eta|\Phi_{SM}|^2|\Phi_H|^2
\end{equation}
The component fields are written as
\begin{equation}
\Phi_{SM}=\frac{1}{\sqrt{2}}\left(%
\begin{array}{c}
  \phi_{SM}+v+iG^0 \\
  G^{\pm}\\
\end{array}%
\right),~~~
 \Phi_H=\frac{1}{\sqrt{2}}(\phi_H+\xi+iG')
\end{equation}
where $v$($\simeq 246\gev$) and $\xi$ are vevs around which the $\Phi_{SM}$
and $\Phi_H$ are expanded. The $G$ fields are Goldstone bosons, which
can be removed from actual calculation by imposing the unitary
gauge. After diagonalizing the mass matrix, we rotate from the
gauge eigenstates ${\phi_{SM},\phi_H}$ to mass eigenstates ${h,H}$.
\begin{eqnarray}
   \phi_{SM}&=&\cos\omega h+\sin\omega H\\
   \phi_H &=&-\sin\omega h+\cos\omega H
   \end{eqnarray}
the mixing angle $\omega$ and the mass eigenvalues are given by
\begin{eqnarray}
\tan\omega&=&\frac{\eta v\xi}{(-\lambda
v^2+\rho\xi^2)+\sqrt{(\lambda v^2-\rho\xi^2)^2+\eta^2v^2\xi^2}}\label{transf1}\\
m^{2}_{h,H}&=&(\lambda v^2+\rho\xi^2)\pm\sqrt{(\lambda
v^2-\rho\xi^2)^2+\eta^2v^2\xi^2}\nonumber
\end{eqnarray}
For simplicity in writing subsequent formula, we
assume that $m_{h}<m_{H}$ and write $c_{\omega}\equiv\cos_{\omega}$, $s_{\omega}\equiv\sin_{\omega}$.

   If $m_{H}>2m_{h}$, the signature of interest, $H\rightarrow hh$
decay, is allowed kinematically. The partial width of
this decay is
\begin{equation}
\Gamma(H\rightarrow hh)=\frac{|\mu|^2}{8\pi
m_{H}}\sqrt{1-\frac{4m_{h}^2}{m_{H}^2}} \label{transf2}
\end{equation}
where $\mu$ is the coupling of the relevant mixing operator in the
Lagrangian $\triangle\mathcal{L}_{mix}=\mu h^2H$.
\begin{equation}
\mu=-\frac{\eta}{2}(\xi
c_{\omega}^3+vs_{\omega}^3)+(\eta-3\lambda)vc_{\omega}^2s_{\omega}+(\eta-3\rho)\xi
c_{\omega}s_{\omega}^2 \label{transf3}
\end{equation}

   Before going to the discussion of the Higgs mass bounds, it
is helpful to do a parameter space analysis for this model. There
are a total of 7 input parameters relevant for most of our later
discussion: $g, \lambda, v, \eta, \rho, \xi, g_V$, where $g$ is
the $SU(2)_L$ gauge coupling, $g_V$ is defined to be the gauge
coupling constant of $U(1)_{hid}$. $g_V$ in general would appear
in the scattering amplitude of the graphs involving the
$U(1)_{hid}$ gauge boson $V$, and therefore play a role in the
discussion of perturbative unitarity (however, in section 3.1, we
will make a reasonable assumption that results in $g_V$
effectively disappearing in all the relevant formulae). Other
possible input parameters that describe the details of the matter
content of the hidden sector itself are uncertain and we do not
include them here (in our work, they are only relevant to the RGE
of $\rho$, where we just introduce two representative parameters
$E$ and $E'$). $g, v$ are already fixed by collider experiments,
with the values $v=246$GeV, $g=0.65$. In order to study the
phenomenology of the model, we construct some output parameters
from these input parameters which are of more physical interest:
$m_W, G_F, m_{h}, s_{\omega}, m_V, G_{F}^{'}, m_{H}, \mu$, where
we define $G_{F}^{'}$ as the Fermi coupling for the $U(1)_{hid}$
defined in the same way as $G_F$ in the SM. We will see in section
3.1 that $G_{F}^{'}$ plays an important role in the unitarity
bounds. The relevant transformations in addition to eqs.
(\ref{transf1})-(\ref{transf3}) are:
\begin{eqnarray}
m_W=\frac{1}{2}gv,~~
m_V=\frac{1}{2}g_V\xi,~~
G_F=\frac{1}{\sqrt{2}v^2},~~
G_{F}^{'}=\frac{1}{\sqrt{2}\xi^2}.
\end{eqnarray}
Now we have determined that the 4 most important unknown input parameters are
$\{\lambda, \rho, \eta, \xi\}$. The inverse transformation from
$\{m_{h}^2, m_{H}^2, s_{\omega}, \mu\}$ to $\{\lambda, \rho,
\eta, \xi\}$ are
 \begin{eqnarray}
\lambda&=&\frac{M^2_{11}}{2v^2}\\
\rho&=&\frac{M^2_{22}}{2v^2s_{\omega}^2}\left[\frac{\cw^3M^2_{12}+3\cw^2\sw
M_{11}^2-2\cw\sw^2M_{12}^2+2\mu
v}{-2\cw^2M_{12}^2+3\cw\sw M_{22}^2+\sw^2M^2_{12}}\right]^2\\
\eta&=&-\frac{M_{12}^2}{\sw
v^2}\left[\frac{\cw^3M^2_{12}+3\cw^2\sw
M_{11}^2-2\cw\sw^2M_{12}^2+2\mu v}{-2\cw^2M_{12}^2+3\cw\sw
M_{22}^2+\sw^2M^2_{12}}\right]\\
\xi&=&\sw v\left[\frac{-2\cw^2M_{12}^2+3\cw\sw
M^2_{22}+\sw^2M^2_{12}}{-cw^3M_{12}^2-3\cw^2\sw
M^2_{11}+2\cw\sw^2M^2_{12}-2\mu v}\right]
\end{eqnarray}
  where
\begin{eqnarray}
M^2_{11}&=&\cw^2m_{h}^2+\sw^2m_{H}^2\\
M^2_{12}&=&\cw\sw(m^2_{H}-m^2_{h})\\
M^2_{22}&=&\sw^2m^2_{h}+\cw^2m^2_{H}
\end{eqnarray}
In Tables~\ref{trans1} and~\ref{hdecay} we provide 6 benchmark
points in parameter space, some of which will be used  in section
4 for collider physics analysis. They all can satisfy the
theoretical bounds as we shall
see in section 3. We list them in Table~\ref{trans1} and Table~\ref{hdecay}.\\

  \begin{table}[ptb]
\par
\begin{center}%
  \begin{tabular}{|c|c|c|c|}
    \hline
     & Point A& Point B & Point C \\
    \hline
     $\sw^2$&        0.40 & 0.31 & 0.1 \\
    $m_{h}$ (GeV) & 143 & 115 & 120 \\
    $m_{H}$ (GeV)& 1100 & 1140 & 1100 \\
    $\Gamma(H\rightarrow hh)$ (GeV)& 14.6 & 4.9 & 10 \\
    $BR(H\rightarrow hh)$& 0.036& 0.015& 0.095 \\
    \hline
      \end{tabular}
\end{center}
\caption{Points illustrating parameters of trans-TeV mass Higgs boson.
Point C is studied in detail in section 4.}%
\label{trans1}%
\end{table}

\begin{table}[ptb]
\par
\begin{center}%
\begin{tabular}{|c|c|c|c|}
    \hline
     & Point 1& Point 2 & Point 3 \\
    \hline
     $\sw^2$&        0.5 & 0.5 & 0.5 \\
    $m_{h}$ (GeV) & 115& 175 & 225  \\
    $m_{H}$ (GeV)& 300 & 500& 500 \\
    $\Gamma(H\rightarrow hh)$ (GeV)& 2.1 & 17 & 17 \\
    $BR(H\rightarrow hh)$& 0.33 & 0.33 &0.33 \\
    \hline
\end{tabular}
  \end{center}
\caption{Points illustrating parameters that allow large branching
fractions of $H\to hh$.  Each of these points are studied in detail
in section 4.}%
\label{hdecay}%
\end{table}

  $\Gamma(H\rightarrow hh)$ for points 1, 2, 3 are obtained based on the assumption
  that the branching ratio BR$(H\rightarrow
  hh)=1/3$ where BR$=\frac{\Gamma(H\rightarrow
  hh)}{\Gamma(H\rightarrow
  hh)+\sw^2\Gamma^{SM}(\m2)}$.
$\Gamma^{SM}(\m2)$ is the well-known SM result, which can be
obtained from the HDECAY program~\cite{Djouadi:1997yw}.

\section{Theoretical Bounds on Higgs Masses of the Model}
\subsection{Perturbative Unitarity Constraints}
\indent

   The possibility of a strongly interacting WW sector or Higgs sector above
   the TeV scale is an interesting alternative to a perturbative, light Higgs boson.
However, this possibility implies the unreliability of
perturbation theory.  Although this is not a fundamental concern, it would
imply a challenge to the successful perturbative description of precision
electroweak data and would have major implications to LHC results.
In order for the perturbative
description of all electroweak interactions to be valid up to a high scale,
the perturbative unitarity constraint would need to be satisfied. This issue has been
carefully studied for the SM Higgs sector\cite{Lee:1977eg}. They obtained an
upper bound on the Higgs mass by imposing the partial-wave unitarity
condition on the tree-level amplitudes of all the relevant
scattering processes in the limit $s\rightarrow\infty$, where $s$
is the center of mass energy. The result is $m^2_{\phi_{SM}}\leq
\frac{4\pi\sqrt{2}}{3G_F}\simeq$ (700 GeV)$^2$. To get this
result,  we apply
a more restrictive condition as in~\cite{gordy}: $|Re
a_J|\leq\frac{1}{2}$, where $a_J$ is the $J^{\rm th}$ partial wave
amplitude. This is also the condition we will apply for our
model.

  We derive the unitarity constraints for our model
by methods analogous to ref.~\cite{Lee:1977eg}. The
addition of one more Higgs and the mixing effects introduce more
relevant processes and more complex expressions. We impose the
unitarity constraints on both the SM sector and the $U(1)_{hid}$
sector. The analysis for the diagrams involving $V$ is very
similar to those involving the $Z$ boson. For simplicity, we assume
that in the hidden sector, $m_V\ll \m2$, as an analogy to the case
in the SM, where $m_W\ll m_H$. With this approximation, $g_V$ will
not appear in the scattering amplitude, only $G_F'$ is relevant.
We list the set of 15 inequalities in the Appendix, and their
corresponding processes. For simplicity, we did not transform them
to purely input or output parameter basis, but kept them in a
mixing form as they were derived for compact expressions. Unlike
the situation in SM, it would be hard to solve this complex set of
inequalities analytically to get the Higgs mass bounds. Instead
using the Monte Carlo method, we generated $60^4\sim10^7$ points
in the input parameter space with basis $\{\lambda, \eta, \rho,
\xi \}$. In order to be consistent with our discussion of
perturbative TeV physics, we liberally set the allowed regions of
these
input parameters to be:
\beq
    \lambda\subset[0,4\pi], \rho\subset[0,4\pi],
    \eta\subset[-4\pi,4\pi], \xi\subset[0,5\tev]
  \eeq
Then we pick out the points that satisfy all 15 inequalities,
and make $m_{H}-m_{h}$ plots for certain narrow ranges of
the mixing angle $\sw^2$ which is an important output parameter for
collider physics study.  The allowed region can be read
from the shape of these plots (obviously, for this multi-dimension
parameter space, the bounds on Higgs mass are
dependent on the mixing angle).

\begin{figure}[h]
\vspace{0cm} \center
\includegraphics[width=12cm]{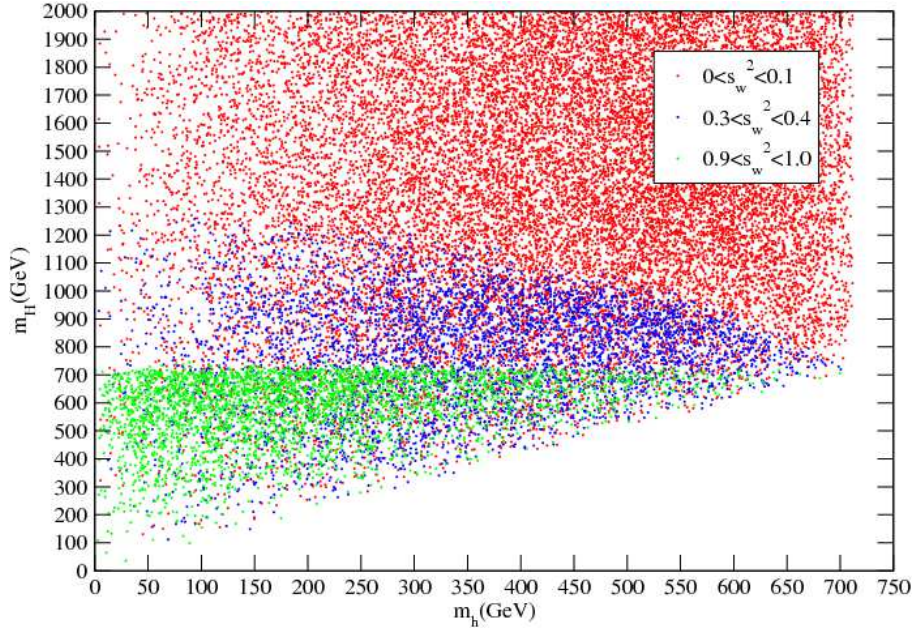}\\
\caption{Scatter plot of solutions in the $m_H$ vs. $m_h$ plane
that satisfies perturbative unitarity constraints.  Separate colors
are used depending on what range $s^2_\omega$ falls within. }
\label{fig1}
\end{figure}

\begin{figure}[h]
\vspace{0cm} \center
\includegraphics[width=12cm]{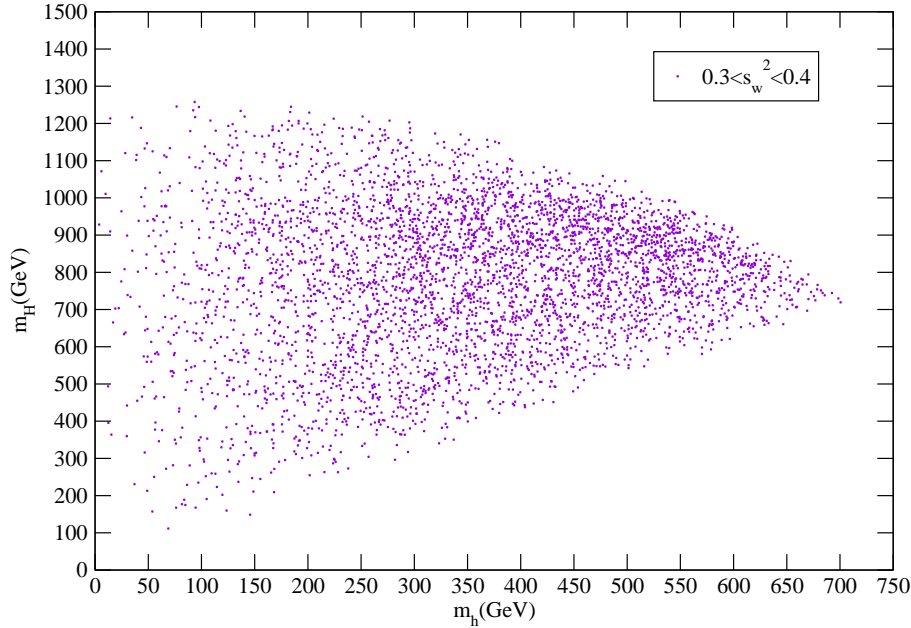}\\
\caption{Scatter plot of solutions in the $m_H$ vs. $m_h$ plane
that satisfies perturbative unitarity constraints.  This plot
is only for points that fall within $0.3<s^2_\omega < 0.4$.}
\label{fig2}
\end{figure}

   Fig.~\ref{fig1} combines the plots for 3 typical mixing
regions -- small mixing, medium mixing and large mixing for
comparison. We can tell that for the lighter physical Higgs
boson mass, the
upper bound always stays the same as the well-known SM
case---around 700 GeV. However, for the heavier Higgs boson in the
spectrum, the bound is loosened: for small mixing it can be as high
as 15 TeV given our parameter ranges (in Fig.~\ref{fig1}, we cut the upper limit at 2 TeV to reduce the
size of the graph as well as improve the presentability of the graph),
for medium mixing can be above 1 TeV --- both are well above the
canonical upper limit of the Higgs boson mass based on unitarity
considerations.  For large mixing limit, the canonical 700 GeV bound
applies for both of the physical Higgs. These observations agree
with our intuition. The intermediate mixing region is of significant
phenomenological interest, since it can not only generate a heavy
Higgs boson --- especially a trans-TeV Higgs which is not well anticipated by
the experiments, yet may be worth attention --- but also can produce
the heavy Higgs boson at a considerable production rate at colliders (we
know that the coupling of $H$ to SM particles is proportional to
$\sw$). That is why we amplify the plot for the medium mixing region
in Fig.~\ref{fig2} to demonstrate the bound shape more clearly. Meanwhile,
the small mixing region can also be interesting, since as $\sw$
decreases, the decay width narrows down which is good for
detection, although the production rate gets lower.

Based on the
considerations described above, we choose 3 typical points from those that
are allowed by all the perturbative unitarity bounds and can generate a
trans-TeV Higgs: points A, B and C, as we
listed in Table 1 at the end of section 2.
They are labelled by the output parameter basis $\{\sw^2, m_{h},
m_{H}, \Gamma(H\rightarrow hh)\}$. Point A and B are from
medium mixing region ($\sw^2=0.40$ for point A is actually the
maximum mixing angle that can allow a $m_{H}$ larger than 1.1
TeV among all the points that satisfy unitarity constraints),
point C is from the small mixing region. We will make precision
electroweak analysis for these 3 points in section 3.3 and study
the collider physics of trans-TeV Higgs bosons in section 4.1.

\subsection{Triviality bounds and Vacuum Stability Bounds}
\indent

  Besides perturbative unitarity, triviality and vacuum stability
are two additional concerns which impose theoretical
constraints on the Higgs mass. Now we want to see if they would put
more stringent bounds on the Higgs mass than those given by unitarity.
In the SM, both of them are actually relevant to the properties of the
parameter $\lambda$ at the high scale, which are analyzed using the RG
equation of $\lambda$. The triviality bound is given based on the
requirement that the Landau pole of $\lambda$ from the low-scale
theory perspective is above the scale
of new physics. The vacuum stability bound is given based on the
requirement that $\lambda$ remains positive up to the scale of new
physics. Now we already can see that the bounds derived from these
two considerations are not definite, as they depend on the scale of new
physics. In the SM, the bounds for the value of $\lambda$ at the
electroweak scale are equivalent to the upper and lower limits for
Higgs boson because of the simple proportion relation
$m_{\Phi_{SM}}^2=2\lambda v^2$, where $v\simeq 246\gev$. As
reviewed in~\cite{Reina:2005ae}, for a $1\tev$ new physics scale,
$160\gev  <m_H<750\gev$. (This is actually a rough estimation based
on 1-loop RGE and without taking into threshold corrections. More
accurate analysis would be subtle.) However, it is easy to tell
that these constraints do not apply for our model where the
physical Higgs spectra are determined by four input parameters
$\lambda, \eta, \rho, \xi$, not just $\lambda$. Therefore, we need
to first derive the RG equations for all these four parameters and
see what we can say for the Higgs mass bounds based on them.

   Here we give the 1-loop results.
For convenience, we suppose that the RGEs run above the
EWSB scale, so that all the masses are zero and we
can safely work with gauge eigenstates. (Actually, as is well
known, the RGEs of dimensionless couplings are independent
of mass parameters, which supports the validity of our assumption.)

  1-loop RGE for $\lambda$ in the SM can be found in \cite{gordy}.
The addition of the hidden sector Higgs boson contributes another term
to the RGE, which results from the mixing term in the Lagrangian
$\frac{1}{4}\eta \phi_H^2\phi_{SM}^2$ ($\phi_{H}$ runs in the loop). The full
result is:
\begin{equation}
\frac{d}{dt}\lambda=\frac{1}{16\pi^2}\left\{\frac{1}{2}\eta^2+12\lambda^2+6\lambda
y_t^2-3y_t^4-\frac{3}{2}\lambda(3g^2+g^2_1)+\frac{3}{16}[2g^4+(g^2+g_1^2)^2]\right\}
\label{lambda}
\end{equation}
where $g_1$ is the gauge coupling of $U(1)_{Y}$, $y_t$ is the top
Yukawa coupling. The first term comes from the interaction between
$\phi_H$ and $\phi_{SM}$.

   For $\rho$, there is also a 1-loop contribution from the graph
where $\phi_{SM}$ runs in the loop. The other terms in the RGE of $\rho$
come from the self-interactions in the hidden sector, e.g. the
coupling between $\phi_H$ and the hidden sector matter---we denote
all these terms by $E$. The result is
\begin{equation}
\frac{d}{dt}\rho=\frac{1}{16\pi^2}(\eta^2+10\rho^2+E) \label{rho}
\end{equation}

   The RGE of $\eta$ involves only two graphs: with $\phi_{SM}$ or $\phi_H$
running in the loop. We eventually get:
\begin{equation}
\frac{d}{dt}\eta=\frac{1}{16\pi^2}\eta \left[6\lambda+
4\rho+2\eta+3y_t^2-\frac{3}{4}(3g^2+g^2_1)+E' \right] \label{eta}
\end{equation}
We can see from eqs.(\ref{lambda})-(\ref{eta}) that the
perturbative properties of $\lambda$, $\rho$ and $\eta$ can be
nice although they are model dependent. However, we can hardly
draw any quantitative conclusions regarding, especially, the Higgs
masses bounds -- they depend on four unknown parameters, the
detailed content of hidden sector matter represented by parameters
$E$ and $E'$, threshold corrections, etc. All of these
uncertainties make the prediction for the triviality and stability
bounds quite model dependent. Meanwhile, such large freedom allows
us to reasonably expect that the points that satisfy the unitarity
conditions are also allowed by triviality and stability
constraints in a large region of full parameter space (with
parameters for hidden sector itself included). A practical
application of this observation is that now we can reasonably
assume that the points from section 3.1 can also pass the test of
triviality and vacuum stability.

\subsection{Contraints from Precision Electroweak Measurements}
\indent

   Precision electroweak measurements also give indirect bounds on
the Higgs boson mass based on the fact that the virtual excitations of
the Higgs boson
can contribute to physical observables, e.g. W boson mass,
considered in precision tests of the SM. For the one doublet Higgs boson
in the SM, precision EW analysis puts a 200 GeV upper limit at
$95\%$ C.L~\cite{lep}. Here we do not plan to make a full analysis
to derive the mass bounds in a general way.
Alternatively, we  focus on the point A, B and C, of which we have
made an $S-T$
analysis to see if they can satisfy the constraints
from experiments. This is actually a way to check for our model
the `existence' of the points allowed by precision EW
measurements.

   The relevant calculations are analogous to those for the SM
Higgs boson. We just need to double the number of involved graphs,
since there are two Higgs bosons now, and put $\sw$ or $\cw$ on some
vertices. The resulting values for $S$ and $T$ for points A, B and
C are consistent with \cite{lep}:
\beq
{\rm
A:}~(S,T)=(0.05,-0.10),~~ {\rm B:}~(S,T)=(0.02,-0.06),~~ {\rm
C:}~(S,T)=(-0.01,-0.01)
\eeq
and
\beq
{\rm 1:}~(S,T)=(0.01,-0.03),~~ {\rm 2:}~(S,T)=(0.05,-0.07),~~
{\rm 3:}~(S,T)=(0.06,-0.09)
\eeq
where we have chosen $m_H=150$ GeV
as the SM reference point where $(S,T)=(0,0)$.
   We compare these results with the $S-T$ contour in \cite{lep} which gives the constraints on $(S,T)$
 from the most recent precision electroweak measurements. Point C is on the boundary of the allowed
region, and therefore satisfies the precision EW constraints. Points 1-3,
A and B seem to be mildly out of the 68\% C.L.\ allowed region. According to
the direction of their shifts relative to the center of the
contour, they have the same effects as a heavy Higgs in the SM.
However, contributions from the unspecified elements of the model -- in particular the
$Z'$ contributions --
can compensate the effect of a heavy Higgs by pulling the $(S,T)$
back towards the center\cite{Peskin:2001rw}. It is easy to tell that such a solution
could also apply to our model by the $Z'$ from its
$U(1)_{hid}$ hidden sector gauge symmetry.

    Therefore, now we can come to the conclusion that all the three interesting points can
satisfy all the known theoretical bounds on Higgs mass under a few
reasonable assumptions. The next step is to send them to the
collider physics analysis so that we can tell whether we can
discover such interesting phenomenology in future experiments.


\section{Large Hadron Collider Studies}

In this section, we consider phenomenological implications for new
physics searches at the LHC. In our framework, we have two Higgs bosons
that are in general mixtures of a SM Higgs boson and a Higgs boson that
carries no charges under the SM gauge groups.  Thus, no state is precisely
a SM Higgs boson and no state is precisely of a singlet nature.  More importantly,
by construction, neither $H$ nor $h$ have full SM Higgs couplings to any
state in the SM.  Production rates are therefore always reduced for $h$ or $H$
compared to the SM Higgs.

Reduced production cross-sections present a challenge for LHC discovery
and study.  Depending on the mass of the SM Higgs boson, there are already
significant difficulties for discovery without the additional worry of reduced couplings.
Nevertheless, opportunities present themselves as well.  For one, the reduced
production cross-section also correlates with a more narrow-width scalar state.
The width of the SM Higgs boson grows so rapidly with its mass (by cubic power)
that by the time its mass is above $\sim 800\gev$ the Higgs boson width is
so large that it begins to lose meaning as a particle. Reduced couplings, and
therefore a reduced width, of a heavy
Higgs boson can bring it into the fold of familiar, narrow-width particles.  We study
this point below to demonstrate that even a Higgs boson with mass greater than
$1\tev$ (i.e., a trans-TeV Higgs boson) can be searched for and found at
the LHC in this scenario.

Another attempt at turning a negative feature into a new angle for searching, is to
accept that two heavily mixed Higgs states could exist, and search for the decay
of the heavier one to the lighter.  These $H\to hh$ decays could be copious enough
that the first discovery of the Higgs boson would be through the simultaneous discovery
of $H$ and $h$ via $H$ production followed by $H\to hh$.  We study this possibility
at the LHC and find that indeed this may be possible.

To begin the discussion, we first state some of the choices we have made to
simulate LHC physics.  We have used
Madgraph \cite{Stelzer:1994ta} to generate all matrix elements. We then use
MadEvent \cite{Maltoni:2002qb}, with the CTEQ6 \cite{Pumplin:2002vw} PDF
set, to generate both signal and background
event samples for all the studies in this paper.  Renormalization and factorization scales
are set to $m_H$ for calculating signal cross-sections.

To partially simulate detector and showering effects, parton energies are smeared by a gaussian function of width
$\sigma/E=0.68/\sqrt{E}\oplus 0.044$ ($E$ is in units of GeV),
from Table 9-1 in~\cite{atlastdr}.   Photon and lepton energies are not smeared.
We assume a $b$-tagging efficiency of $50\%$ and mistag
rates for $c$,$g$, and $uds$ partons of
$10\%$, $1.5\%$, and $0.5\%$, respectively.  All jets are required to have
 $p_T >$ 30 GeV and $\left \vert \eta \right \vert <$ 4.5, where $\eta$ here refers
 to the pseudo-rapidity ($\eta=-\ln\tan(\theta/2)$ with $\theta$ being the polar angle
 with respect to the beam).  Leptons and photons
 are required to be separated from jets by $\Delta$R$>$0.4 and from one another
 by $\Delta$R$>$0.2, where $\Delta R=\sqrt{(\Delta\eta)^2+(\Delta\phi)^2}$ ($\phi$ is
 the azimuthal angle).
 Jets must be separated from each other by $\Delta$R$>$0.7, or they
 are merged.
We do not apply any triggering or reconstruction efficiencies.

\subsection{Narrow Trans-TeV Higgs boson}

Earlier we showed that a very heavy Higgs boson can be compatible with all
known constraints.  Its couplings will necessarily be less than those of the
SM Higgs boson, but if it is mixed with the SM Higgs boson, the mass eigenstate
$H$ can be searched for and discovered even if its mass is above $1\tev$.
We show here that a very narrow resonance, which is implied by the reduced
couplings,
may enable background normalizations to be determined using
sideband techniques which are not possible with the very large widths for
heavy SM Higgs bosons.

As we do not consider decays to new particles, the final state topologies
are the same as the searches investigated for 1 TeV Higgs bosons (see \cite{AtlasTeVHiggs}),
though the cross-sections and width are both reduced
by sin$^2\omega$ compared to a SM Higgs of the same mass.
We set sin$^2\omega$ = 0.1 and $M_H$ = 1.1 TeV (see point C of Table~\ref{trans1}).
This leads to a
width $\Gamma_H$=95 GeV and NLO cross-section $\sigma_H$= 7.1 fb for vector boson fusion.
The comparison SM values, which we augment to compute our decay widths and cross-section,
are obtained from HDECAY \cite{Djouadi:1997yw} and \cite{Han:1992hr}.


We begin with a study of $qqH$ production followed by
$H\to WW\to \ell\nu jj$.
The significant difference between previous SM studies \cite{AtlasTeVHiggs} and our study is that
the reduced Higgs width allows for reducing systematic uncertainties in the measurement of
background rates.
We do not do a complete set of background calculations, but instead argue, based on the simulations
we have done, that the normalizations for all backgrounds can be determined
from mass reconstruction distributions.

We require one lepton ($e$,$\mu$) with $p_T > $ 100 GeV, $\left \vert \eta \right \vert  < $2.0 and
missing energy transverse to the beam $\slashET > $ 100 GeV.  We also
require two ``tagging'' jets with $\left \vert \eta \right \vert  > $ 2.0.  Finally, we require the two
highest $p_T$ jets to have $p_T > $ 100 GeV and reconstruct to within
20 GeV of the $W$ mass.  We relax the separation cut between these two jets
to $\Delta$R$>$0.3.  (Reconstructing highly-boosted, hadronic $W$ bosons
has been studied \cite{AtlasBoostedW}.)

The $WWjj$ background is calculated with $\mu_F$=$\mu_R$=$M_W$.
The $W+$4j background has not
been simulated, but is not expected to have a kinematic shape which would
complicate determining its normalization from data.
The $t\bar{t}jj$ background is calculated with both scales set to $M_{top}$.
We simulate $t\bar{t}jj$ such that the two jets from the
production stage are explicitly the two tagging jets used in the analysis.  While this is not
a complete description of the $t\bar{t}+ n$ jet background, we wish only to make the point
that there are no kinematic features that would complicate deriving
its normalization from data.  A more complete background analysis implies
that full reconstruction and showering will not overwhelm the signal, as shown
in ref.~\cite{AtlasTeVHiggs}.

\begin{figure} \center
\includegraphics[angle=90]{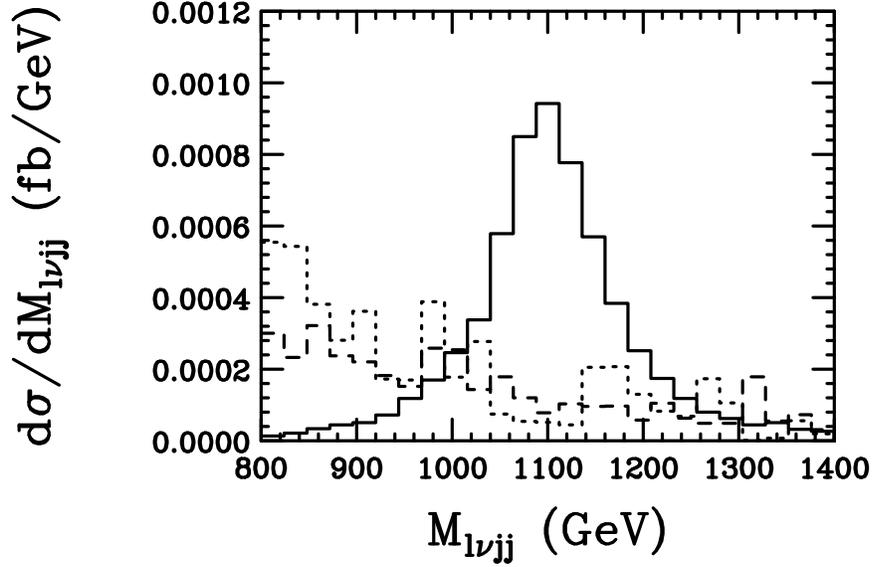}
\caption[transverse mass]{Differential
cross-section as a function of the invariant mass of the $\ell$, $\slashET$ and
two jets reconstructing to the $W$ mass
for $H \rightarrow WW \rightarrow \ell\nu jj$ (solid), $WWjj$ (dashed),
and $t\bar{t}jj$ (dotted).  }
\label{mass_lvjj}
\end{figure}

Fig.~\ref{mass_lvjj} shows the differential cross-section as a function
of the invariant mass of the lepton, $\slashET$ and two highest $p_T$ jets.  Below 900 GeV,
the distribution is almost entirely background, allowing for an extraction of the
$W$ and $t\bar{t}$  normalizations.
As the figure demonstrates, one can rather easily distinguish the trans-TeV
Higgs boson from the background after all the cuts once there is enough data for
the distribution to be filled.  As expected, luminosity is critical. In this case, after all
cuts, the integral of the signal from $1.0\tev < M_{l\nu jj}<1.3\tev$ yields 12.8 events in
$100\xfb^{-1}$, while the total background amounts to $7.7$ events.  For a more assured
discovery and more accuracy on the Higgs boson mass, one would need
more data.  Nevertheless, this signal channel alone demonstrates the plausibility of
discovering a Higgs boson in the trans-TeV mass region. Analysis of more decay channels,
if these tantalizing results emerged, would further increase the significance and accuracy
of discovery.

%


For example, a heavy Higgs boson that decays to $WW$ with a sizeable branching fraction will
also decay to $ZZ$, which can be used to increase the significance of the discovery
and test the self-consistency of the theory.  In this case we
look at decays to two $Z$ bosons which then decay to either
$\ell \ell jj$ or $\ell \ell \nu \nu$.  A mass reconstruction for the first case would
yield a distribution similar in shape to Fig.~\ref{mass_lvjj}, so we instead
plot the transverse mass distribution for $\ell \ell \nu \nu$.  This final state
has the virtue of only one significant background ($ZZjj$) which is under
better theoretical control that the $Z$+$\geq$4j background.  Still,
$ZZ\rightarrow\ell \ell jj$ has a larger rate, though a potentially large
background from $ZZ$+$\geq$4j production, and should be considered as well.

We require same-flavor, opposite-sign leptons, each with $p_T > $ 100 GeV and
$\left \vert \eta \right \vert <$ 2.0 which reconstruct to within 5 GeV of the $Z$ mass.
We also require
two tagging jets with $\left \vert \eta \right \vert  > $ 2.0 and
$\slashET>$100 GeV.
The only significant SM background is from $ZZjj$ production.  We calculate this
background at LO using factorization and renormalization scales set to $M_Z$.

\begin{figure} \center
\includegraphics[angle=90]{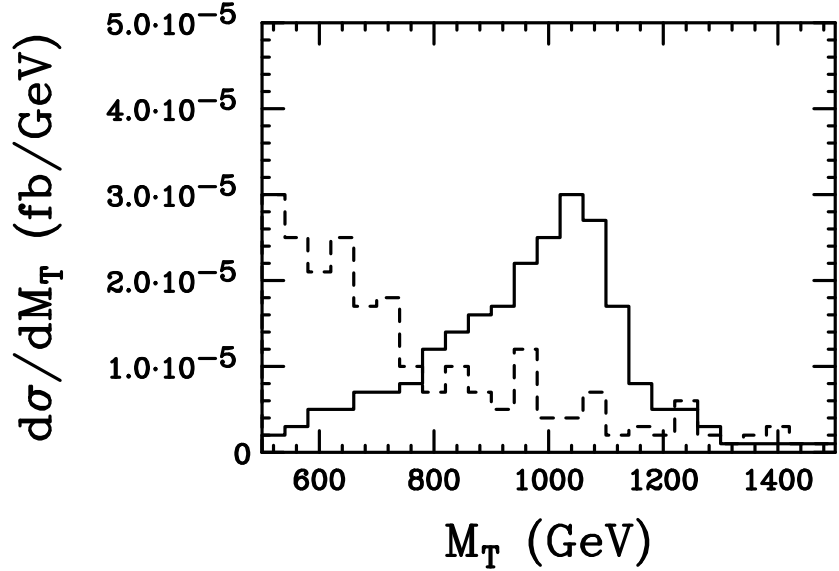}
\caption[transverse mass]{Differential
cross-section as a function of transverse mass of the $Z$ and $\slashET$
for $H \rightarrow ZZ \rightarrow \ell\ell\nu\nu$ (solid) and
the $ZZjj$ background (dashed).  }
\label{transversemass}
\end{figure}

Fig.~\ref{transversemass} shows the differential cross-section as a function
of the transverse mass $M_T$, where $M_T^2=2 \left \vert  p_{T_{\ell \ell}} \right \vert $$\left \vert \slashET \right \vert (1-\cos\phi)$
and $\phi$ is the angle between the reconstructed leptonic $Z$ and the $\slashET$
 in the transverse
plane.  The production cross-section and branching ratios are small enough in this model
that this channel is not as important without large amounts of data,
but the relatively small backgrounds and
distinctive shape imply that it could be important for other models.

Fig.~\ref{transversemass} demonstrates that the transverse mass variable is a good
discriminator of signal to background as long as enough integrated luminosity is
obtained at the collider. The combination of this channel (and several others
similar to it) with the
$H\to WW$ results of the previous section increases the significance of discovery.
In this particular example final state, there are 3.9 signal events compared to 1.4 background events
in the transverse mass region $0.8\tev < M_T<1.4\tev$ with $500\xfb^{-1}$.
Discovering
hidden sector Higgs theories with reasonable significance by adding up all possible
channels\footnote{There are many more channels to exploit, potentially including the $ZZ$ channel
arising from $gg\to H$ production. This could be a productive channel since tagging jets
are not needed to reduce the $t\bar tX$ background.}
will come first at
lower luminosity, but the results above indicate that
careful checks of various final states and self-consistency are possible, albeit at
a much higher luminosity stage of the collider.  This would give us the opportunity to study
the precise nature of the trans-TeV Higgs boson through its various branching fractions.

\subsection{$H\rightarrow hh$ Signal}

We now examine Higgs-to-Higgs decays, and consider whether these
decays might be the first evidence for either the $H$ or $h$ boson~\cite{Hhh studies} at the LHC.
Although
it might be possible to effectively search for both Higgs bosons when the heavier
one is in the trans-TeV mass range, we focus on somewhat lighter Higgs boson
masses in this section which clearly show the feasibility of this kind of search over
much of parameter space.

We normalize $gg\rightarrow H$ production
to the NNLO rates \cite{Catani:2003zt} of 10.3 pb and 5.7 pb
for 300 GeV and 500 GeV SM Higgs bosons, respectively.
VBF production is normalized to the NLO rates \cite{MCFM}
of 1.3 pb and 0.54 pb for 300 GeV  and 500 GeV SM Higgs bosons, respectively.
Both cross-sections
are then multiplied by sin$^2\theta$=0.5 to obtain the production rates for $H$ and $h$.


To begin with, let us suppose that the heavy and light Higgs mass eigenstates
are $m_H= 300\gev$ and $m_h = 115\gev$, respectively (see point 1 of Table~\ref{hdecay}).
Even if the $115\gev$
mass eigenstate had full-strength SM couplings, its discovery is by no means easy.
A SM Higgs with mass around 115 GeV
relies principally on the $t\bar{t}h\to t\bar t b\bar b$ production
channel as well
as direct production $gg\rightarrow h \rightarrow \gamma\gamma$.
If signal production is reduced by half (i.e., $\sin^2\omega=1/2$)
and/or background rates are greater than calculated,
or systematic uncertainties
prove larger than anticipated, the discovery of this lighter Higgs boson
will require significantly more
data.  We consider the possibility that the lighter
higgs may be discovered instead through $H\rightarrow hh\rightarrow \gamma \gamma b \bar{b}$
decays.  In our example point, as with all example points in this section, the branching ratio
of $H\to hh$ is $1/3$.



\begin{figure} \center
\includegraphics[angle=90]{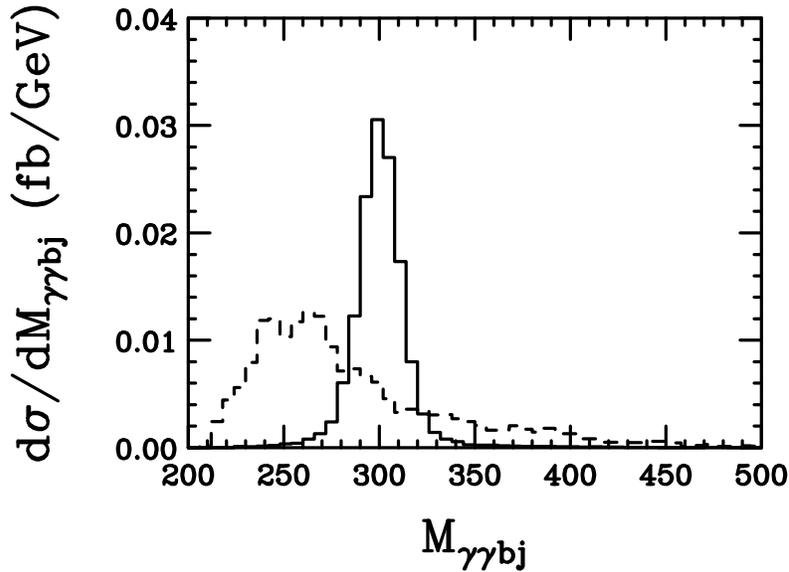}
\caption[myybb]{Differential
cross-section as a function of invariant mass of $\gamma \gamma b \bar{b}$
for $H\rightarrow hh \rightarrow \gamma\gamma\bar{b} {b}$ (solid) and
the sum of the backgrounds (dashed) requiring one $b$-tag. }
\label{invmass}
\end{figure}

To determine the viability for discovery, we first calculate the background
processes that could contribute to $\gamma\gamma b\bar b$ events in the SM.
The factorization and renormalization scales ($\mu_F$ and $\mu_R$) used for computing this
background are set to the leading $p_T$ jet
in the event.  The observable we define requires two photons and two jets, with at least one
jet tagged as containing a $b$ quark.
We furthermore require $\left\vert m_h-m_{\gamma\gamma}\right\vert <$ 2 GeV,
$\left\vert m_h-m_{j_1j_2}\right\vert <$ 20 GeV, and $\left\vert m_H-m_{\gamma\gamma j_1 j_2}\right\vert <$ 20 GeV.
Fig.~\ref{invmass} shows the reconstructed invariant mass of the two photons and
two jets with one b-tag.

The general strategy to extract the signal over SM background is the same as for the
supersymmetric $H\rightarrow hh \rightarrow \gamma \gamma b\bar{b}$
search channel \cite{Richter-Was:1996ak}.  We argue here that this signature
is important for a broad range of models.  Although it is only considered important
for supersymmetry scenarios with small
tan $\beta$, this decay channel looks to be important for a wide
range of parameter space for Higgs-mixing scenarios because of its relatively
high rate of triggering and narrow mass reconstruction.

As the numbers in Table~\ref{nums1} indicate, we find that signal-to-background ratios
for both the single and double tag samples are sufficient for discovery.  Even
after detector triggering and reconstruction efficiencies are applied, there should
still be enough events for a discovery in the first few years of data taking at the LHC.
We thus argue that, for this model, the light Higgs might be discovered through these
$H \rightarrow hh \rightarrow \gamma \gamma b \bar{b}$ decays before appearing
in the more conventional
$t\bar{t}h$, $qq\rightarrow qqh$, or $gg \rightarrow h$ searches, especially if the
systematics for those channels prove to be more challenging than expected.


\begin{table}[ptb]
\par
\begin{center}%
\begin{tabular}
[c]{|c|c|c|}\hline
Channel & 1 tag & 2 tags \\\hline
$H\rightarrow hh$             & 24       &  12     \\\hline\hline
$\gamma \gamma b b$    & 0.4    &  0.2        \\\hline
$\gamma \gamma b c$    & 0.15    & 0.01   \\\hline
$\gamma \gamma b j$     & 1         & 0.009        \\\hline
$\gamma \gamma c c$    & 1.2      & 0.069     \\\hline
$\gamma \gamma c j$     & 3.6      & 0.042   \\\hline
$\gamma \gamma j j$     & 1.8       & 0.007         \\\hline\hline
Total background             &  8.2     &  0.34         \\\hline

\end{tabular}
\end{center}
\caption{Numbers of ``$\gamma\gamma b\bar b$" (defined in the text)
events for 30 fb$^{-1}$ after applying all cuts with 1 or 2 $b$-tags required.
Summation of charge conjugation is implied (e.g. $b$=$b$+$\bar{b}$) and $j$=$u,d,s$.
The Higgs boson properties
are those of point 1 in Table~2.}%
\label{nums1}%
\end{table}




If the lighter Higgs boson is above the $2m_W$ threshold, qualitatively
new features of the signal develop that we explore now.  For example, let us
suppose that the lighter
Higgs boson is $175\gev$ and that
the heavier Higgs boson mass is $m_H=500\gev$, which allows $H\to hh$ decays
with $1/3$ branching fraction (see point 2 of Table~\ref{hdecay}).  For this point
we again have a reduction of $1/2$ in the cross-section due to $s_\omega^2=1/2$.


In this case, the most common final state
for $H\rightarrow hh$ decays will be four $W$ bosons.  This signature
has been studied in the context of dihiggs production \cite{Baur:2002qd}
but SM dihiggs production is on the order of 10-30fb \cite{Dawson:1998py}.
In Higgs-mixing scenarios, $H\rightarrow hh$ production is generically an order of
magnitude or two larger.

We divide the study up into two searches by the number of leptons in the final state.
First,
we require three leptons, where the opposite-sign pairs must have opposite flavor (OSOF).  This
follows the strategy in \cite{Baur:2002qd} for reducing the large $Z$/$\gamma W^{\pm}$
background.  We also look at events with four leptons and demand that opposite-sign,
same-flavor lepton pairs not reconstruct to within 5 GeV of $M_Z$.
One could also use same-sign (SS) dilepton searches, but the backgrounds are significantly larger and more difficult to predict so we do not explore this here.


$W^{\pm}W^{\pm}W^{\mp}$ samples
are all generated at $\mu_F=\mu_R=M_W$.
The $t\bar{t}Z$ and $t\bar{t}W$ samples are generated with $\mu_F=\mu_R=M_{top}=$ 175 GeV.
All backgrounds are generated at LO and no K-factors are applied.
A  $\slashET>$ 50 GeV cut has been applied to all searches.  Leptons that do not
satisfy $p_T >$ 20 GeV and $\left\vert \eta \right\vert <$ 2.0, or are not isolated
from other leptons or jets, are considered lost.  $Z$/$\gamma W^{\pm}$ with
 $Z$/$\gamma \rightarrow \tau \bar{\tau}$ has been investigated for the OSOF 3$\ell$ and
 found to be small.

\begin{table}[ptb]
\par
\begin{center}%
\begin{tabular}
[c]{|c|c|c|c|}\hline
Channel                                                      & $\sigma$ (fb)  & OSOF 3$\ell$  \\\hline
$H\rightarrow hh\rightarrow WWWW$   &  920                       &        56                    \\\hline\hline
$W^{\pm}W^{\pm}W^{\mp}$                     &  109                        &          5                        \\\hline
$t\bar{t}Z$                                                    &  580                        &          1                          \\\hline
$t\bar{t}W^{\pm}$                                       &  740                         &         15                          \\\hline
\end{tabular}
\end{center}
\caption{Numbers of 3$\ell$ OSOF events for 30 fb$^{-1}$.  The Higgs boson properties
are those of point 2 in Table~2.}%
\label{leptonswwww}%
\end{table}

Table \ref{leptonswwww} shows the number of  OSOF 3$\ell$ events expected for 30 fb$^{-1}$.
The dominant $t\bar{t}W$ background
may have large NLO corrections, but applying a $b$-jet veto would further
reduce it by approximately 64$\%$, while reducing the
signal by only a few percent.  Additionally, there are 8 four-lepton events which could be
used.

For comparison, in this model we expect 9 $H \rightarrow ZZ \rightarrow$ 4$\ell$ events for 30 fb$^{-1}$
 satisfying the lepton cuts described above, with each opposite-sign, same-flavor pair reconstructing
 to within 5 GeV of $M_Z$,
 and satisfying $\left\vert M_H-m_{4\ell}\right\vert < \Gamma_H$, where $\Gamma_H$=51 GeV.
For the same cuts, the irreducible $pp\rightarrow ZZ$ background yields 8 events, using
$\mu_F=\mu_R=M_Z$ and applying no K-factors.

Based on the numbers in Table \ref{leptonswwww}, we argue that, for this model, the
heavier Higgs can be discovered through the
$H \rightarrow hh \rightarrow $ OSOF 3 $\ell$ channel in the first few years
at the LHC.  Furthermore, this channel may compete with more conventional
searches, such as $H \rightarrow ZZ \rightarrow$4$\ell$ for an early discovery.

Finally, we comment on the situation of point 3 in Table~\ref{hdecay} where
the lighter Higgs is heavier than $2M_Z$. In this case, the branching ratios to $WWZZ$ and $ZZZZ$ can
be significant.  For example, using the same parameters as above,
if the mass of the lighter Higgs is raised to 225 GeV, the cross-sections
for $H\rightarrow hh \rightarrow WWZZ$ and $H\rightarrow hh \rightarrow ZZZZ$ are
425 fb and 87 fb.  The $WWWW$ final state is still the largest branching ratio, but
other searches involving $Z$ boson final states would aid discovery.



\section{Conclusions}

The Large Hadron Collider holds much promise for discovering new
particles and interactions.  Many ideas of physics beyond the SM
that explain electroweak symmetry breaking involve states that are
coupled directly to the Standard Model gauge bosons. For example,
supersymmetry, technicolor and extra dimensions all have exotic states
that are direct participants in the electroweak story.  However, there
are states that do not couple to the SM gauge bosons that may
contribute to understanding the full picture of EWSB (e.g., singlet states
that get vevs to produce the $\mu$ term in supersymmetry) or help
solve other problems not directly connected to electroweak physics
(e.g., singlets breaking exotic gauge groups in string-inspired theories).

In this article, we have investigated a renormalizable interaction between
the SM Higgs boson and a Higgs boson of a hidden sector.  This gives us
one of the most incisive methods to probe the existence of states that
have no SM gauge charges.  The phenomenological challenge to this scenario
is that all couplings of the mixed Higgs bosons are less than the would-be
SM couplings for a SM Higgs boson of the same mass.  However, small
compensating advantages were exploited here: a reduced coupling means
a reduced width, which turns a trans-TeV Higgs boson into a definable
narrow-width state to search for, and the existence of two Higgs bosons enables
us to search for decays of the heavier Higgs boson to the lighter one.
In both cases, we were able to study examples from the parameter space of
discovery.  We therefore like to emphasize the importance of doing
searching for a Higgs boson in standard channels well into the trans-TeV
mass region.  We also like to reemphasize, from the point of view of these
hidden sector ideas, that there is a potential opportunity to discover both a
heavy Higgs boson and a light Higgs boson through $H\to hh$ decays.
This is an especially attractive channel to exploit in the circumstance that
a light $h$ boson is particularly hard to find due to reduced production cross-section
which is generically predicted in these theories.

\bigskip
\noindent {\bf Acknowledgements:} We wish to thank G. Kane, F.
Maltoni, D.Morrissey and D. Rainwater for discussions. This work
has been supported in part by the Department of Energy and the
Michigan Center for Theoretical Physics (MCTP).


\section *{Appendix: Unitarity Inequalities}

The 15 relevant processes  that give
non-vanishing constant amplitudes when $s\rightarrow\infty$ (with
$m_W,m_V\ll m_H,m_h$) are \\

1. $W^{+}_LW^{-}_L\rightarrow W^{+}_LW^{-}_L$ ($s$-, $t$-channels) \\

2. $Z_LZ_L\rightarrow Z_LZ_L$ ($s$-, $t$-, $u$-channels)\\

3. $Z_LZ_L\rightarrow W^{+}_LW^{-}_L$ (only the $s$-channel Higgs
exchange is relevant)\\

4. $HH\rightarrow HH$ (only contact graphs are relevant), in mass eigenstates, including:\\
   (4.1) $h h\rightarrow h h$\\
   (4.2) $h h\rightarrow h H$\\
   (4.2) $h h\rightarrow H H$\\
   (4.4) $H H\rightarrow h H$\\
   (4.5) $H H\rightarrow H H$\\

5. $HH\rightarrow W^{+}_LW^{-}_L/Z_LZ_L$ ($t$-,$u$- channel gauge
boson exchange and s-channel Higgs exchange are all relevant),
including:\\
   (5.1) $h h\rightarrow W^{+}_LW^{-}_L/Z_LZ_L$\\
   (5.2) $h H\rightarrow W^{+}_LW^{-}_L/Z_LZ_L$\\
   (5.3) $H H\rightarrow W^{+}_LW^{-}_L/Z_LZ_L$\\

6. $V_LV_L\rightarrow V_LV_L$ ($s$-, $t$-, $u$-channels) \\

7. $HH\rightarrow V_LV_L$ ($t$-,$u$- channel gauge boson exchange and
s-channel Higgs exchange are all relevant), including:\\
   (7.1) $h h\rightarrow V_LV_L$\\
   (7.2) $h H\rightarrow V_LV_L$\\
   (7.3) $H H\rightarrow V_LV_L$\\

The corresponding conditions derived from those 15 processes are
listed below in order:
\begin{eqnarray}
&&\frac{G_F(\cos^2\omega m^2_{h}+\sin^2\omega
m^2_{H})}{4\sqrt{2}\pi}\leq\frac{1}{2}\\
&&\frac{3G_F(\cos^2\omega m^2_{h}+\sin^2\omega
m^2_{H})}{8\sqrt{2}\pi}\leq\frac{1}{2}\\
&&\frac{G_F(\cos^2\omega m^2_{h}+\sin^2\omega
m^2_{H})}{8\sqrt{2}\pi}\leq\frac{1}{2}\\
&&\left|\frac{3}{8\pi}(\lambda\cos^4\omega+\rho\sin^4\omega+\eta\sin^2\omega\cos^2\omega)\right|\leq\frac{1}{2}\\
&&\left|\frac{3}{8\pi}[-\lambda\cos^3\omega\sin\omega+\rho\sin^3\omega\cos\omega-\frac{1}{2}\eta(-\sin\omega\cos^3\omega+\cos\omega\sin^3\omega)]\right|\leq\frac{1}{2}\\
&&\left|\frac{1}{4\pi}[-\frac{3}{2}\lambda\sin^2\omega\cos^2\omega-\frac{3}{2}\rho\sin^2\omega\cos^2\omega-\frac{1}{4}\eta(\sin^4\omega+\cos^4\omega-4\sin^2\omega\cos^2\omega)]\right|\leq\frac{1}{2}\\
&&\left|\frac{3}{8\pi}[-\lambda\sin^3\omega\cos\omega+\rho\cos^3\omega\sin\omega-\frac{1}{2}\eta(-\cos\omega\sin^3\omega+\sin\omega\cos^3\omega)]\right|\leq\frac{1}{2}\\
&&\left|\frac{3}{8\pi}(\lambda\sin^4\omega+\rho\cos^4\omega+\eta\cos^2\omega\sin^2\omega)\right|\leq\frac{1}{2}\\
&&\left|\frac{1}{16\pi}\{-4\sqrt{2}G_Fc_\omega^2m^{2}_{h}-\frac{6c_\omega}{v}(-\lambda
vc_\omega^3+\rho\xi s_\omega^3-\frac{1}{2}\eta vc_\omega
s_\omega^2+\frac{1}{2}\eta\xi s_\omega
c_\omega^2)\right. \\\nonumber&&\left. -\frac{2s_\omega}{v}[-3\lambda
vc_\omega^2s_\omega-3\rho\xi s_\omega^2c_\omega-\frac{1}{2}\eta
v(-2s_\omega c^2_\omega+s_\omega^3)-\frac{1}{2}\eta\xi(-2c_\omega
s^2_\omega+c_\omega^3)]\}\right|\leq\frac{1}{2}\\
&&\left|\frac{1}{16\pi}\{-2\sqrt2G_Fs_\omega
c_\omega(m^2_{h}+m^2_{H})-\frac{2c_\omega}{v}[-3\lambda
vc_\omega^2s_\omega-3\rho\xi s_\omega^2c_\omega-\frac{1}{2}\eta
v(-2s_\omega
c^2_\omega+s_\omega^3)\right.\\\nonumber&&\left. -\frac{1}{2}\eta\xi(-2c_\omega
s^2_\omega+c_\omega^3)]-\frac{2s_\omega}{v}[-3\lambda
vs_\omega^2c_\omega+3\rho\xi c_\omega^2s_\omega-\frac{1}{2}\eta
v(-2c_\omega s^2_\omega+c_\omega^3)-\frac{1}{2}\eta\xi(2s_\omega
c^2_\omega-s_\omega^3)]\}\right|\leq\frac{1}{2}\\
&&\left|\frac{1}{16\pi}\{-4\sqrt{2}G_Fs_\omega^2m^{2}_{H}-\frac{6s_\omega}{v}(-\lambda
vs_\omega^3-\rho\xi c_\omega^3-\frac{1}{2}\eta vs_\omega
c_\omega^2-\frac{1}{2}\eta\xi c_\omega
s_\omega^2)\right. \\\nonumber&&\left. -\frac{2c_\omega}{v}[-3\lambda
vs_\omega^2c_\omega+3\rho\xi c_\omega^2s_\omega-\frac{1}{2}\eta
v(-2c_\omega s^2_\omega+c_\omega^3)-\frac{1}{2}\eta\xi(2s_\omega
c^2_\omega-s_\omega^3)]\}\right|\leq\frac{1}{2}\\
&&\frac{3G^{'}_F(\sin^2\omega m^2_{h}+\cos^2\omega
m^2_{H})}{8\sqrt{2}\pi}\leq\frac{1}{2}\\
&&\left|\frac{1}{16\pi}\{-4\sqrt{2}G^{'}_Fs_\omega^2m^{2}_{h}+\frac{6s_\omega}{\xi}(-\lambda
vc_\omega^3+\rho\xi s_\omega^3-\frac{1}{2}\eta vc_\omega
s_\omega^2+\frac{1}{2}\eta\xi s_\omega
c_\omega^2)\right. \\\nonumber&& \left. -\frac{2c_\omega}{\xi}[-3\lambda
vc_\omega^2s_\omega-3\rho\xi s_\omega^2c_\omega-\frac{1}{2}\eta
v(-2s_\omega c^2_\omega+s_\omega^3)-\frac{1}{2}\eta\xi(-2c_\omega
s^2_\omega+c_\omega^3)]\}\right|\leq\frac{1}{2}\\
&&\left|\frac{1}{16\pi}\{+2\sqrt2G^{'}_Fs_\omega
c_\omega(m^2_{h}+m^2_{H})+\frac{2s_\omega}{\xi}[-3\lambda
vc_\omega^2s_\omega-3\rho\xi s_\omega^2c_\omega-\frac{1}{2}\eta
v(-2s_\omega
c^2_\omega+s_\omega^3)\right. \\\nonumber&& \left. -\frac{1}{2}\eta\xi(-2c_\omega
s^2_\omega+c_\omega^3)]-\frac{2c_\omega}{\xi}[-3\lambda
vs_\omega^2c_\omega+3\rho\xi c_\omega^2s_\omega-\frac{1}{2}\eta
v(-2c_\omega s^2_\omega+c_\omega^3)-\frac{1}{2}\eta\xi(2s_\omega
c^2_\omega-s_\omega^3)]\}\right|\leq\frac{1}{2}\\
&&\left|\frac{1}{16\pi}\{-4\sqrt{2}G_F^{'}c_\omega^2m^{2}_{H}-\frac{6c_\omega}{\xi}(-\lambda
vs_\omega^3-\rho\xi c_\omega^3-\frac{1}{2}\eta vs_\omega
c_\omega^2-\frac{1}{2}\eta\xi c_\omega
s_\omega^2)\right. \\\nonumber&& \left. +\frac{2s_\omega}{\xi}[-3\lambda
vs_\omega^2c_\omega+3\rho\xi c_\omega^2s_\omega-\frac{1}{2}\eta
v(-2c_\omega s^2_\omega+c_\omega^3)-\frac{1}{2}\eta\xi(2s_\omega
c^2_\omega-s_\omega^3)]\}\right|\leq\frac{1}{2}
\end{eqnarray}


\end{document}